\documentstyle[aasms4,12pt,epsf]{article}

\begin{document}

\title{ Near Infrared Observations of a Redshift 5.34 Galaxy: Further
Evidence for Dust Absorption in the Early Universe\altaffilmark{1}}

\altaffiltext{1}{Based on observations obtained at the W.M. Keck
Observatory, which is operated jointly by the California Institute of
Technology and the University of California}

\author{L. Armus\altaffilmark{2,3},  K. Matthews\altaffilmark{3}, G.
Neugebauer\altaffilmark{3}, B.T. Soifer\altaffilmark{3}}

\altaffiltext{2}{SIRTF Science Center, 100-22, Caltech, Pasadena, CA
91125}

\altaffiltext{3}{Palomar Observatory, 320-47, Caltech, Pasadena, CA
91125}


\begin{abstract}

\gdef\Msun{M_{\odot}}

Imaging at 1.25$\mu$m and 2.20$\mu$m has been obtained of the field
containing the galaxy (RD1) found at redshift $z=5.34$ by Dey, et
al.(1998).  This galaxy has been detected at 1.25$\mu$m, while the
lower redshift ($z=4.02$) galaxy also found in the same field by Dey,
et al. was detected at both 1.25$\mu$m and 2.20$\mu$m.  Comparison to
stellar population synthesis models indicates that if RD1 is a young
(t $<10^{8}$ yrs) galaxy, significant reddening (A$_{V}>0.5$ mag) is
indicated.  Combined with observations of other high redshift
systems, these data show that dust is likely to be an important
component of young galaxies even at redshifts of $z > 5$.  The
extinction-corrected monochromatic luminosity of RD1 at 1500\AA\ is
then a factor of about three larger than L$_{1500}^*$ as determined
by Dickinson (1998) for $z\sim3$ starburst galaxies.  The implied
star formation rate in RD1, corrected for extinction, is $\sim
50-100{\it h}_{50}^{-2}{\it M}_{\odot}$yr$^{-1}$.

\medskip

\end{abstract}


\section{Introduction}
\label{sec:introduction}

In the last several years, the combination of new wavelength dropout
discovery techniques (e.g. Steidel et al. 1996) coupled with the deep
imaging power of the Hubble Space Telescope and the spectroscopic
capabilities of a new generation of large ground-based telescopes, has
lead to a blossoming of the study of galaxies at redshifts of $z>2$.  A
recent breakthrough in this field has been the discovery of a galaxy at a
redshift of $z=5.34$ by Dey et al. (1998).  Besides being the highest
redshift object presently known, this galaxy is important because it is
apparently neither gravitationally lensed, nor the host of a powerful
active nucleus.  Thus there is hope that detailed studies of this system
and others like it at $z\sim5$ will shed light on the intrinsic, stellar
properties of very young galaxies.

Because visible observations of objects at such high redshifts are
sampling the restframe ultraviolet part of the spectrum, they are very
sensitive to the presence of even small amounts of intervening dust.  Dust
is important at these redshifts for three simple reasons.  First,
intrinsic luminosities and star formation rates are always underestimated
when dust is not taken into account.  This may have vast implications for
metal production at early epochs (Madau et al.  1996, Meurer et al.
1997).  Second, dust may hide entire populations of high redshift,
star-forming galaxies from detection because of sample selection effects
(e.g. Dey, Spinrad \& Dickinson 1995).  Third, dusty interstellar media in
galaxies at $z>3-4$, by their very nature, imply previous episodes of star
formation and enrichment, pushing the first epoch of star formation to
even higher redshifts.  Although ``pollution" of the ISM can happen
relatively quickly from a truly primordial state (e.g.  Matteucci \&
Padovani 1993), dusty systems at high redshift always provide a lower
limit to the redshift of the earliest star formation.  Since there is
mounting evidence for dust in high redshift galaxies (e.g.  Dunlop et al.
1994, van Ojik et al. 1994, Ivison 1995, Dey, Spinrad \& Dickinson 1995,
Knopp \& Chambers 1997, Sawicki \& Yee 1998) it is natural to ask whether
dust is present in galaxies at $z>5$.

To understand the true nature of the $z=5.34$ galaxy discovered by Dey, et
al. (1998), hereafter referred to as RD1 - for Red Dropout 1, it is
necessary to know whether it is being viewed through a significant amount
of dust.  To achieve this, it is essential that such objects be observed at
the longest possible wavelengths.  We report in this paper serendipitous
observations made of this galaxy at J (1.25$\mu$m) and K (2.20 $\mu$m).
The galaxy is clearly detected at J, and simple fits to the observed
optical-infrared colors suggest a reddened, young stellar population.  In
the discussion we adopt the cosmological parameters H$_{\rm0}=50$ km
s$^{-1}$ Mpc$^{-1}$ and q$_{\rm 0}=$0.5.  Throughout this paper we refer
to broad-band magnitudes measured with respect to Vega, such that J$=0$
mag and K$=0$ mag correspond to 1578 Jy and 646 Jy, respectively.

\smallskip

\section{Observations and Data Reduction}
\label{sec:observations}

As part of an observational study of high redshift radio galaxies (Armus,
et al. 1998a, 1998b) we observed the most distant known radio galaxy,
6C0140+32 in August 1997 and January 1998 using the near infrared camera
(NIRC) on the K1 telescope of the W.M. Keck Observatory.  The instrument
is described in detail by Matthews and Soifer (1994).  It has a
256$\times$256 InSb array with 0.15$\arcsec \times$0.15$\arcsec$ pixels
for a 38$\arcsec \times$ 38$\arcsec$ field of view.  On 18 August 1997,
the field was observed at K under conditions of thin cirrus and good
seeing ($\sim0.5\arcsec$).  On January 16, 1998 we observed this field
under clear, photometric conditions at J and K in excellent seeing
($0.4-0.5''$ FWHM).  On 17 Jan 98 we also observed the field at K under
conditions of good seeing with thin cirrus.

The target for the observations was centered in the field of view, and
observations were made with individual integration times of 60 seconds.
In the August observations the target was moved in a regular 9 point
square grid covering a size 10$\arcsec$ on a side, with the center of the
grid moved by  $1 - 2 \arcsec$ between grids. In the January observations,
the target was moved randomly over an area of 10$\arcsec \times$
10$\arcsec$ between individual observations in order to affect optimal sky
subtraction.

Because the galaxy RD1 and the companion galaxy BD3 (for Blue Dropout 3)
at a redshift $z=4.022$ were only $\sim 12 \arcsec$ from the targeted
object, they were contained in virtually every frame, and the combined
data achieved the full sensitivity of the observation at the location of
RD1 and BD3 at each wavelength.

A bright star in the center of the NIRC field was used to accurately
register the successive frames for coaddition.  To remove time-variable
fluctuations in illumination, separate sky and normalized flat-field
frames were created from the data for each set of three images, by taking
the median of the nearest 7-9 frames.  After being trimmed to a size of
$251 \times 251$ pixels, the individual data frames were sky subtracted,
flat fielded, and shifted to a common dc level after known bad pixels were
flagged.  These processed images were then aligned, using integer pixel
shifts, and combined using a clipped mean algorithm.  The FWHM of a point
source in the final mosaic, as determined from several stellar images in
the field, was approximately 0.5$\arcsec$ at J and 0.45$\arcsec$ at K.
The photometry was based soley on the data from 16 Jan 98, when conditions
were photometric.  The data were calibrated by reference to the infrared
standard stars of Persson et al. (1998).

\section{Results}

The J and K images of the field centered on RD1 are shown in Figure 1.
The J data are from 16 Jan 98 only, while the K data are from 18 Aug 97,
16 Jan and 17 Jan 98. We obtained a total of 70 and 92 minutes of
integration at J and K, respectively.

In order to bring out the faint, high redshift sources RD1 and BD3, the
data in Figure 1 have been convolved with a circular gaussian having a
FWHM equal to the measured stellar FWHM in the final J and K-band
mosaics.  In this figure, $12\arcsec$ sections of the mosaics are
displayed.  Objects RD1 and BD3 are both detected in the J-band image.  At
K, BD3 is marginally detected, while RD1 is undetected.  The J-band image
of RD1 is resolved in the north-south direction, while that of BD3 is
unresolved.

Because the objects RD1 and BD 3 are separated by $\sim 1.2\arcsec$
obtaining photometry of both systems required special care.  A square
beam, $0.75\arcsec$ on a side, was centered on both RD1 and BD3.  The sky
was taken to be adjacent and to the south for RD1, and adjacent and to the
north for BD3.  As a check on this small beam photometry, aperture
photometry was obtained of the combined flux from both objects, and this
total was distributed between the two objects based on the one dimensional
distribution of flux along the line between the two sources.  The two
methods agreed well, and the magnitudes are reported in Table 1.  For RD1
we measure J$=24.3\pm0.3$ mag and K$>23.1$ mag ($3\sigma$ limit).  For BD3
we measure J$=24.1\pm0.3$ mag and K$=23.0\pm0.4$ mag.  In Table 1 we have
also included the I-band magnitudes of these objects as reported by Dey et
al. (1998).

To use the broadband magnitudes to derive continuum fluxes requires
correction for contamination by strong emission lines contained within the
filter bandpasses.  In the case of RD1 the J filter contains redshifted
CIII] 1909\AA\ (at $1.206\mu$m), while the K filter contains redshifted
[OII]3727\AA\ (at $2.355\mu$m).  Heckman et al (1998) have recently
compiled a composite UV spectrum of nearby starburst galaxies that shows
the CIII] emission-line equivalent width to be generally less than
$10$\AA.  This is small compared to the uncertainty in the measured J-band
magnitude of RD1.  Similarly, the average radio galaxy spectrum in
McCarthy (1993) can be used as an upper limit on the strength of CIII] in
RD1.  If the equivalent width of CIII] in RD1 is the same as in a typical
radio galaxy (32\AA), the CIII] line would contribute only about 8\% of
the total flux in the J filter.  Thus we conclude that the correction for
line emission in the J-band for RD1 is not significant.  However, unlike
CIII], the [OII] 3727\AA\ line can be quite strong in star forming
galaxies, having rest-frame equivalent widths of $50-100$ \AA\ (e.g. Cowie
et al. 1995, Gallagher et al. 1989). At an equivalent width of 100\AA\,
[OII] would contribute 14\% of the ($3\sigma$) limit we measure for the
K-band flux from RD1.

In the case of BD3, there are no strong lines expected to be present in
the J filter, unless MgII at 2800\AA\ has a broad, blue wing,  which is
unlikely for a starburst galaxy.  The only emission feature that could
contribute to the K-band measurement is the H$\beta$ line.  While
potentially strong, the redshifted wavelength of H$\beta$ (2.44$\mu$m)
places it at the very edge of the K filter and it should have a negligible
effect on the total flux.

\smallskip

\section{Discussion}
\label{sec: Discussion}

The most important result of these observations is that RD 1 was detected
at J.  This immediately implies that it is a much redder system than
expected for any dust free, young galaxy model.  The expected color of an
unreddened, star forming galaxy with ${\it f}_{\nu}\sim$const is
I$-$J$\sim0.5$ mag and I$-$K$\sim1.5$ mag, while the apparent color of RD1
is I$-$J$=2.2\pm0.3$ mag and I$-$K $< 3.5$ mag.  Note, that source BD3 is
also redder than expected for an unreddened starforming galaxy, having
I$-$J$=1.2\pm0.3$ mag and I$-$K$=2.2\pm0.5$ mag.

For comparison, the colors of the galaxy CL1358+62G1 at a redshift of
$z=4.92$ (Franx et al. 1997, Soifer et al. 1998) are I$-$J$=1.2\pm0.1$ mag
and I$-$K$=2.1\pm0.2$ mag.  For CL1358+62G1, Soifer et al. have shown that
galaxy models with reddenings of $0.1 <$ E(B$-$V) $< 0.4$ mag provide
substantially better fits to the overall energy distribution than do
models with no reddening.

Similarly, Sawicki \& Yee (1998) find that the rest-frame UV-optical
colors of spectroscopically confirmed $z > 2$ Lyman break galaxies in the
HDF imply significant internal dust obscuration at these redshifts.
Again, young populations with E(B$-$V) $\sim0.3$ provide better fits to
the data than do older, un-reddened models.  Although RD1 and BD3 appear
redder than the six, $z > 3$ sources in Sawicki \& Yee, there is
considerable scatter among the HDF galaxies.  The range in I$-$J color of
the HDF galaxies (closest to the J$-$K colors of RD1 and BD3 in rest
wavelength given the smaller redshifts of the Sawicki \& Yee galaxies) is
$-0.98 <$I$-$J$< +0.65$ mag, while the J$-$K colors of RD1 and BD3 are $<
+1.2$ mag and $+1.1$ mag, respectively.  Since we have argued above that
the J-band magnitude of RD1 is not likely to be significantly contaminated
by strong line emission, it seems that the measured continuum colors of
RD1 and BD3 are likely to be affected by dust.

With only an I$-$J color, and a limit on the I$-$K color, it is difficult
to disentangle the effects of intrinsic galaxy color and reddening for
RD1.  However, we can identify a range of allowable ages and reddening
values for a set of simple galaxy models by using the synthetic spectral
energy distributions of Bruzual \& Charlot (1993, 1996).  Because we are
simply trying to set the constraints on reddening and stellar age, we
choose to fit aging, instantaneous starburst models, since these are the
intrinsically reddest galaxies at a given age, and thus they require the
least amount of dust reddening to match a given observed spectrum.
Continuous star formation models will always require more reddening by
dust.

Before comparing the data to the models, a correction must be made to the
observed I-band flux density for the presence of foreground absorption.
Dey et al. (1998) correct their measured I-band magnitude for Ly$\alpha$
emission and estimate I$=26.5 \pm0.1$ mag in a 1.5'' diameter beam (see
Table 1).  However there is no light detected blueward of Ly$\alpha$ in
the spectrum, and much of this absorption may occur at redshifts well
below $z=5.3$.  Thus, the intrinsic rest-frame UV flux density of RD1 is
underestimated if the Ly$\alpha$ emission-line alone is taken into
account.  By comparing the Keck I-band filter and CCD responses to the RD1
spectrum, we have estimated that this correction amounts to approximately
34\% of the measured I-band light, assuming all of the light blueward of
Ly$\alpha$ is absorbed in intervening material.  Once this is taken into
account, RD1 has an I-band magnitude of $\sim 26.2$ mag.  It is this value
that we use in all subsequent modelling.

The comparison of the present data with the Bruzual \& Charlot models
shows that for RD1, either young, dusty models, or "old'', dust-free
models provide acceptable fits to the I, J, and K-band data.  We have used
both solar and 0.2Z$_{\odot}$ metallicity models, although we consider the
latter more representative of the galaxies at these redshifts.  Both sets
of models require significant absorption, A$_{V}>0.5$ mag, for stellar
populations with ages significantly less than $10^{8}$ yrs.  At an age of
$10^{8}$ yrs, the 0.2Z$_{\odot}$ model requires A$_{V}=0.3$ mag, while the
solar metallicity model requires no reddening to fit the data.  Both sets
of models deliver unacceptable fits to the data for ages of
$5\times10^{8}$ yrs or greater, regardless of the reddening.  In all
cases, the SMC reddening curve of Gordon \& Clayton (1998) and the Bruzual
\& Charlot models with a Salpeter IMF have been used.

If RD1 is indeed a reddened, young galaxy then both its luminosity and
star formation rate are larger than those estimated by Dey et al.(1998).
If  RD1 has an age of significantly less than $10^8$ yr and a metal
abundance $\sim0.2$Z$_{\odot}$, the visual extinction is larger than
$\sim0.5$ mag, and the extinction at 1270\AA\ is a minimum of $\sim 2.7$
magnitudes, based on the reddening curve for the SMC (Gordon and Clayton
1998).  Assuming the reddening to the Ly$\alpha$ line is the same as that
to the far UV continuum, the Ly$\alpha$ and UV continuum luminosity are
larger by factors of $10-15$ than those determined by Dey et al. (1998)
for no reddening.  The monochromatic luminosity of RD1 at 1500\AA\ is then
a factor of $\sim3$ larger than L$_{1500}^*$ determined by Dickinson
(1998) for $z\sim3$ starburst galaxies.  The star formation rate in RD1 is
then $\sim 50-100{\it h}_{50}^{-2}{\it M}_{\odot}$yr$^{-1}$, which is
comparable to the extinction-corrected values estimated by Sawicki \& Yee
for the $z >2$ Lyman break galaxies in the HDF.

\bigskip
\bigskip

\acknowledgments 

We thank R. Goodrich for assistance with the observations, and S.E.
Persson for providing photometric standards in advance of publication.
Discussions with Daniela Calzetti, Arjun Dey, Tim Heckman and David Hogg
are greatly appreciated.  Mike Pahre provided us with the Keck I-band
filter and CCD response curves, and we thank him for that.  We also would
like to thank the referee, Hy Spinrad, for a critical reading of the
manuscript and a number of valuable suggestions.  The W.M. Keck
Observatory is operated as a scientific partnership between the California
Institute of Technology, the University of California and the National
Aeronautics and Space Administration. It was made possible by the generous
financial support of the W.M. Keck Foundation.  Infrared astronomy at
Caltech is supported by grants from the NSF and NASA.  This research has
made use of the NASA/IPAC Extragalactic Database which is operated by the
Jet Propulsion Laboratory, Caltech, under contract with NASA.

\clearpage
\begin{table}


\caption{Photometry of RD1 and BD3}

\smallskip
\begin{tabular}{c c c c c c c }   
\tableline\tableline

Object & I & J & K \\
&  mag & mag & mag  \\
\tableline

RD1 &    26.5$\pm0.1$\tablenotemark{a,b} &   24.3$\pm$0.3 & 23.1\tablenotemark{c}\\

BD3 &  25.1$\pm0.1$\tablenotemark{a} &   24.1$\pm$0.3 & 23.0$\pm$0.4\\

\tablenotetext {a} {I-band magnitudes are from Dey et al. (1998) as measured in a $1.5''$ diameter circular beam.}
\tablenotetext {b} {The I-band magnitude of RD1 has been corrected for Ly$\alpha$ emission in the bandpass as per Dey et al.}

\tablenotetext{c} { 3 sigma limit, corresponding to a 1 sigma limit of 24.3 mag}

\end{tabular}

\end{table}


\clearpage

\begin{center}
{\bf Figure Captions}
\end{center}

\figcaption[fig1.ps]{Images of the field of 6C0140+32 RD1 at J and K
obtained with the W.M. Keck Telescope.  The scale and orientation of each
field is shown.  The images show the locations of the objects RD1
($z=5.34$) and BD3 ($z=4.02$) identified by Dey et al. (1998).  After
coadding the data in each filter, the images have been smoothed with a
gaussian matching the average FWHM of a stellar source in the final mosaic
(see text for details).  The positions of RD1 and BD3 are 10.2'' west,
5.6'' north and 10.6'' west, 6.5'' north of star A ($\alpha =$01:43:43.67,
$\delta =$+32:53:54.3 J2000), respectively, as given in Dey et al.}


\clearpage

\thebibliography{}

\bibitem{} Armus, L., Soifer, B.T., Murphy, T.W. Jr., Neugebauer, G.,
Evans, A.S., and Matthews, K. 1998a, \apj, 495, 276.

\bibitem{} Armus, L., Soifer, B.T., Murphy, T.W. Jr., Neugebauer, G.,
Evans, A.S., and  Matthews, K. 1998b, \apjl, in preparation.

\bibitem{} Bruzual, G. and Charlot, S. 1993, \apj, 405, 538.

\bibitem{} Bruzual, G. and Charlot, S. 1996, GISSEL Population Synthesis
Models.

\bibitem{} Calzetti, D., Kinney, A.L. and Storchi-Bergmann, T. 1994, \apj,
429, 582.

\bibitem{} Cowie, L.L., Hu, E., \& Songalia, A. 1995, \nat, 377, 603.

\bibitem{} Dey, A., Spinrad, H., \& Dickinson, M. 1995, \apj, 440, 515.

\bibitem{} Dey, A.J., Spinrad, H., Stern, D., Graham, J.R., and Chaffee,
F.H. 1998, \apjl, in press.

\bibitem{} Dickinson, M. 1998, {\it The Hubble Deep Field, M. Livio, S.M.
Fall, \& P. Madau eds.}

\bibitem{} Dunlop, J. S. Hughes, D. H., Rawlings, S., Eales, S., \&
Ward, M. J.
 1994, \nat, 370, 347

\bibitem{} Franx, M., Illingworth, G.D., Kelson, D.D., van Dokkum, P.G.
and Tran, K-V. 1997, \apjl, 486, L75.

\bibitem{} Gallagher, J.S., Bushouse, H. and Hunter, D.A. 1989, \aj, 97, 700.

\bibitem{} Gordon, K.D. and Clayton, G.C. 1998, \apj, in press.

\bibitem{} Heckman, T.M., Robert, C., Leitherer, C., Garnett, D.R., \& van
der Rydt, F.1998, preprint.

\bibitem{} Ivison, R.J. 1995, \mnras, 275, L33.

\bibitem{} Knopp, G.P. \& Chambers, K.C. 1997, \apj, 487, 644.

\bibitem{} Leitherer, C. and Heckman, T.M. 1995, \apjs, 95, 9.

\bibitem{} McCarthy, P.J. 1993, \araa, 31, 639.

\bibitem{} Madau, P., Ferguson, H.C., Dickinson, M.E., Giavalisco, M.,
Steidel, C.C., \& Fruchter, A. 1996, \mnras, 283, 1388.

\bibitem{} Matteucci, F., \& Padovani, P. 1993, \apj, 419, 485.

\bibitem{} Matthews, K. and Soifer, B.T. 1994, {\it Infrared Astronomy
with Arrays: the Next Generation, I. McLean ed.} (Dordrecht: Kluwer
Academic Publishers), p.239.

\bibitem{} Meurer, G.R., Heckman, T.M., Lehnert, M.D., Leitherer, C., \&
Lowenthal, J. 1997, \aj, 114, 54.

\bibitem{} Persson, S.E., Murphy, D.C., Krzeminski, W., Roth, M., \&
Rieke, M. 1998, \aj, submitted.

\bibitem{} Sawicki, M. \& Yee, Yee, H.K.C. 1998, \apj, 115, 1329.

\bibitem{} Soifer, B.T., Neugebauer, G., Franx, M., Matthews, K. and
Illingworth, G.D. 1998, \apj Letters, in press.

\bibitem{} Steidel, C.C., Giavalisco, M., Pettini, M., Dickinson, M. and
Adelberger, K.L. 1996a, \apjl, 462, L17.

\bibitem{} van Ojik, R., Rottgering, H.J.A., Miley, G.K., Bremer,
M.N., Macchetto, F., \& Chambers, K.C. 1994, \aap, 289, 54.
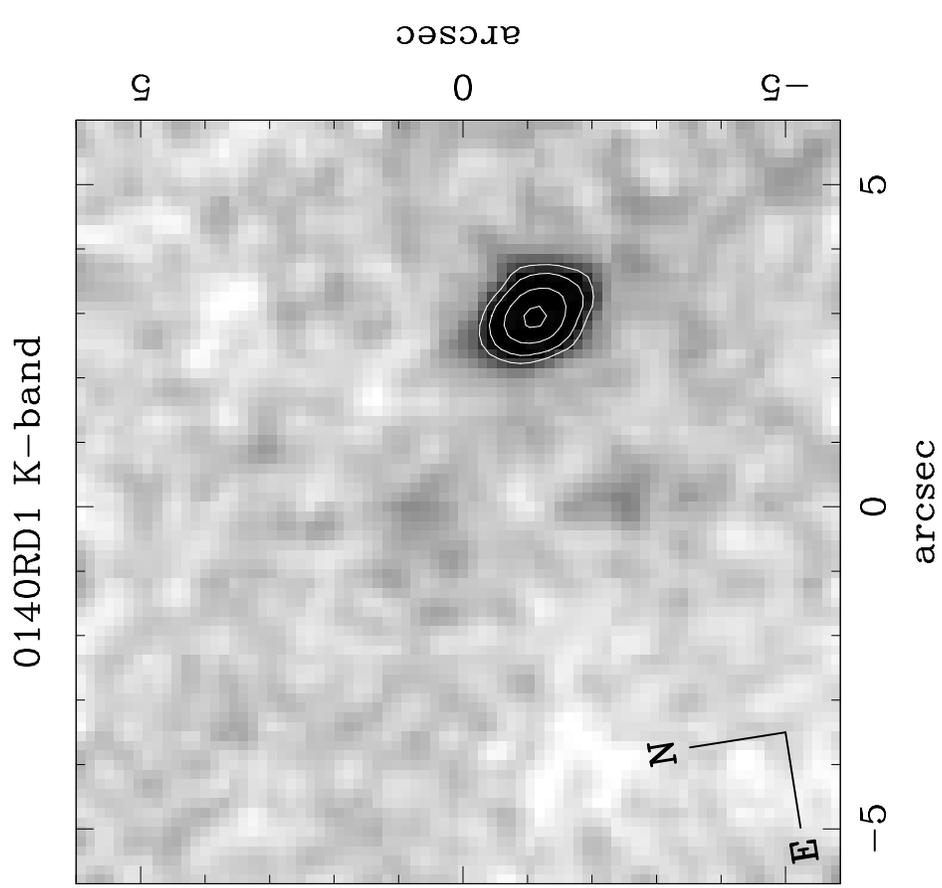
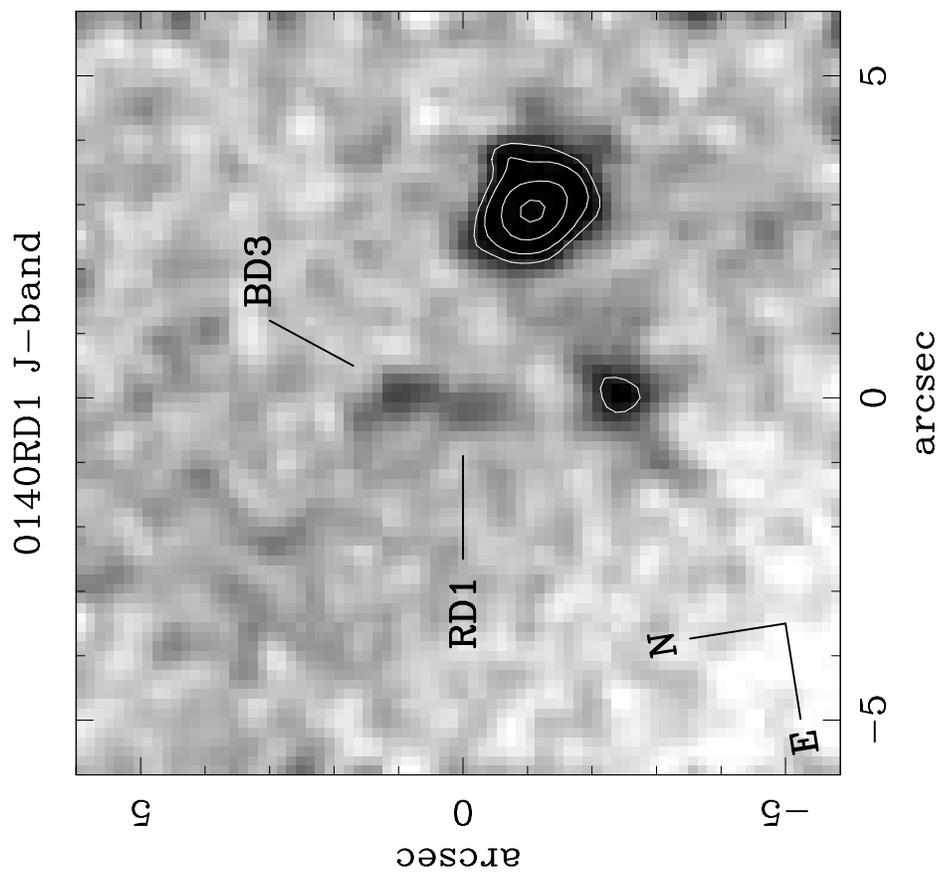
\end{document}